\documentclass[10pt, preprint]{emulateapj}		

\usepackage{graphicx}
\usepackage{color}

\newcommand*{\unit}[1]{\ensuremath{\mathrm{\, #1}}}

\newcommand*{\pcmc}{\unit{cm}^{-3}}	

\newcommand*{\beq}{\begin{equation}}
\newcommand*{\eeq}{\end{equation}}

\newcommand*{\beqa}{\begin{eqnarray}}
\newcommand*{\eeqa}{\end{eqnarray}}

\usepackage{natbib}
\begin{document}

\title{SDO/AIA Detection of Solar Prominence Formation within a Coronal Cavity}


\author{Thomas E.~Berger\altaffilmark{1}, Wei Liu\altaffilmark{2}$^,$\altaffilmark{3}, 
   and B.~C. Low\altaffilmark{4}}	

\altaffiltext{1}{National Solar Observatory, 950 N. Cherry Avenue, Tucson, AZ 85719}
\altaffiltext{2}{Lockheed Martin Solar and Astrophysics Laboratory, B/252 3251 Hanover St., Palo Alto, CA 94304}
\altaffiltext{3}{W.~W.~Hansen Experimental Physics Laboratory, Stanford University, Stanford, CA 94305}
\altaffiltext{4}{High Altitude Observatory, National Center for Atmospheric Research, P.O. Box 3000, Boulder, CO 80307}

\shorttitle{Prominence Formation in Coronal Cavity}
\shortauthors{Berger et al.}
\slugcomment{Accepted by ApJ Letters, September 17, 2012}

\setcounter{footnote}{0}

\begin{abstract}	
We report the first analyses of SDO/AIA observations of  the formation of a quiescent polar crown prominence in a coronal cavity. The \ion{He}{2} 304~\AA\ ($\log\mathrm{T_{max}}\sim 4.8$~K) data show both the gradual disappearance of the prominence due to vertical drainage and lateral transport of plasma followed by the formation of a new prominence 12 hours later. The formation is preceded by the appearance of a bright emission ``cloud'' in the central region of the coronal cavity. The  peak brightness of the cloud progressively shifts in time from the \ion{Fe}{14} 211~\AA\  channel, through the \ion{Fe}{12} 193~\AA\  channel, to the \ion{Fe}{9} 171~\AA\  channel ($\log\mathrm{T_{max}}\sim 6.2, 6.1, 5.8$~K, respectively) while simultaneously decreasing in altitude. Filter ratio analysis estimates the initial temperature of the cloud to be approximately $\log\mathrm{T}\sim 6.25$~K with evidence of cooling over time. The subsequent growth of the prominence is accompanied by darkening of the cavity in the 211~\AA\ channel. The observations imply prominence formation via \emph{in situ} condensation of hot plasma from the coronal cavity, in support of our previously proposed process of magneto-thermal convection in coronal magnetic flux ropes. 
\end{abstract}



\section{Introduction}
\label{sect_intro}

Quiescent prominences, or filaments, are dynamic formations of relatively cool and dense plasma (T $\sim 10^4$~K, $\mathrm{n_{e}}\sim 10^{11}\pcmc$) extending into the much hotter and rarefied corona (T $\sim 10^6$~K, $\mathrm{n_{e}}\sim 10^{8}\pcmc$) over magnetic polarity boundaries far from active regions \citep{Martin:1998,TandbergHanssen:1995ud,Priest:1988,Schmieder:1984ul}. In the ``polar crown'' regions they are frequently observed in so-called ``coronal cavities,'' dark elliptical structures seen during eclipses or in extreme ultraviolet (EUV) and X-ray images of the corona \citep{Kucera:2012,Reeves:2012bu,Schmit:2011ix,Habbal:2010,Fuller:2009jm,Gibson:2006kr}.  Coronal cavities and prominences are the progenitors of coronal mass ejections (CMEs), with the overlying arcade loop system, the cavity, and the prominence comprising the leading dense shell, dark void, and bright trailing material, respectively \citep{Dere:1999ki,Gibson:1998ua,Illing:1986gq}.  Understanding the mechanisms of formation, evolution, and eventual eruption of this complex magnetic system is a central goal of solar physics.

Prominence formation mechanisms can be generally classified into chromospheric transport and coronal condensation \citep{Labrosse:2010bt,Mackay:2010fp}.  In the former, \emph{cool} plasma is lifted from the chromosphere into the corona by siphon flows \citep{PikelNer:1971ft}, magnetic reconnection \citep{Litvinenko:1999gc}, or magnetic flux emergence \citep{Okamoto:2008kg}.  In the latter, \emph{hot} plasma 	
 in the local corona condenses into a prominence \citep{Xia:2012,Luna:2012eb,Karpen:2008hc,Antiochos:1999ef,An:1985fa,1983SoPh...88..219P}. Confirming observations of any of these mechanisms have thus far been elusive. 

Here we report the first analysis of {\it Solar Dynamics Observatory} Atmospheric Imaging Assembly \citep[SDO/AIA;][]{Lemen:2011} observations of  apparent \emph{in situ} formation of a quiescent prominence within a coronal cavity. The results are consistent with earlier studies of cloud prominence formation \citep{Liu:2012gq} and support the hypothesis that coronal cavities and prominences are not in magnetostatic equilibrium, but instead harbor a novel magneto-thermal convection \citep{Berger:2011eoa} with hot upflows and condensation downflows enabling the build-up of magnetic energy and helicity and culminating in an eruptive CME \citep{Zhang:2005dn}.  Recent theoretical analyses \citep{Low:2012tu,Low:2012tv} suggest that MHD condensation has a resistive origin due to runaway thermal collapse of condensing plasmas and spontaneous current sheets \citep{Parker:1994cs,Parker:1953}, leading to rapid condensation and cross-field transport of prominence plasma.

\section{Temporal Sequence of Events}
\label{sect_temporal}

\begin{figure*}
	\includegraphics[width=\textwidth]{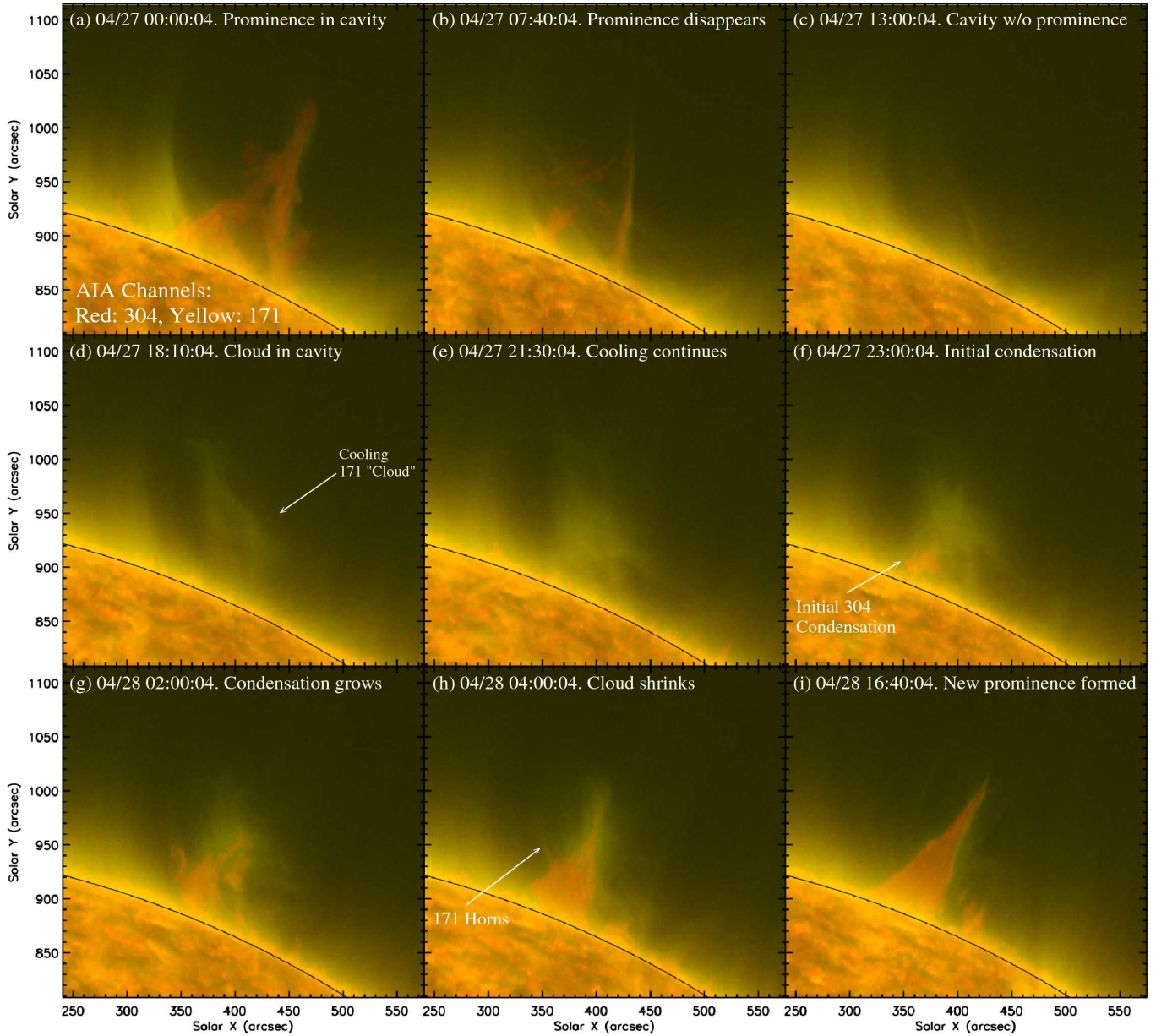} 
	\caption{Time sequence of SDO/AIA images in the 171~\AA\ (yellow) and 304~\AA\ (red) channels. Note the disappearance of the prominence from the cavity, (a)--(c), the appearance of a bright emission cloud, (d)--(e), and the formation and growth of a new prominence, (f)--(i). See also Movies 1a \& 1b, 304~\AA\ overlaid on 171 and 211~\AA, respectively.}
	\label{fig:171+304}
\end{figure*}

The event under study occurred on 27--28 April 2012 on the NW limb (heliographic x,y: +375\arcsec, +900\arcsec\ from disk center). 
Beginning at 04/27 00:00 UT, a tall quiescent polar crown prominence was observed in the AIA \ion{He}{2} 304~\AA\ channel, extending approximately 150\arcsec\  (108 Mm) above the limb.  The prominence was surrounded by a large well-defined coronal cavity that shows maximum contrast in the \ion{Fe}{14} 211~\AA\ channel. The prominence gradually disappears from the 304~\AA\ channel over the course of the next 13 hours, due primarily to downward mass flows, as well as perhaps thermal ``Disparation Brusque'' \citep[DBt;][]{OfmanMouradian:1996}. We note that the time scale for the disappearance is much shorter than required for solar rotation to displace the prominence longitudinally from the limb. Also, STEREO/EUVI data establish that the cavity has a very large longitudinal extent, with prominence plasma remaining at other longitudes. However those same data show that the prominence we observe at the AIA limb is distinct from these other formations. Figures~\ref{fig:171+304}(a)--(c) show this temporal evolution in the AIA 304 and 171~\AA\ channels.   

Following the disappearance, the cavity region near the limb remains prominence-free for some 5--6 hours. Meanwhile, a ``cloud'' of hot plasma gradually accumulates near the core of the cavity. The cloud has the highest contrast against the background in the \ion{Fe}{9} 171~\AA\ channel and develops into its brightest apparition by 04/28 01:28~UT. At about 04/27 22:00~UT there is an initial appearance of  304~\AA\ emission at the lower edge of the cloud. Figures~\ref{fig:171+304}(d)--(f) illustrate this initial cloud formation and prominence appearance sequence. 

Figures~\ref{fig:171+304}(g)--(i) show that the 304~\AA\ emission continues to expand in both height and width over the next several hours. Simultaneously, the size of the 171~\AA\ emission cloud decreases markedly. By 04/28 04:00~UT the 171~\AA\ emission is confined largely to the region surrounding the growing prominence.	
Thin threads of 171~\AA\ emission can be seen curving outward and upward from the prominence at this time. Such prominence ``horns'' have been observed in both prior AIA data \citep{Regnier:2011cr} as well as SOHO data \citep{Vourlidas:2012tz,Plunkett:2000ud}. In the final period of the observation, the 171~\AA\ emission is confined to a very narrow ``Prominence Corona Transition Region'' (PCTR) sheath \citep{Parenti:2007ba} with some residual projections upwards into the cavity. At this point, the coronal cavity is again well-defined in the 211~\AA\  channel. The total time from initial re-appearance in the 304~\AA\ channel to the fully extended prominence in Panel~(i) is approximately 18 hours.

\section{Multispectral Analysis of the Cooling Cloud}
\label{sect_cloud}

\begin{figure}
	\includegraphics[width=0.5\textwidth]{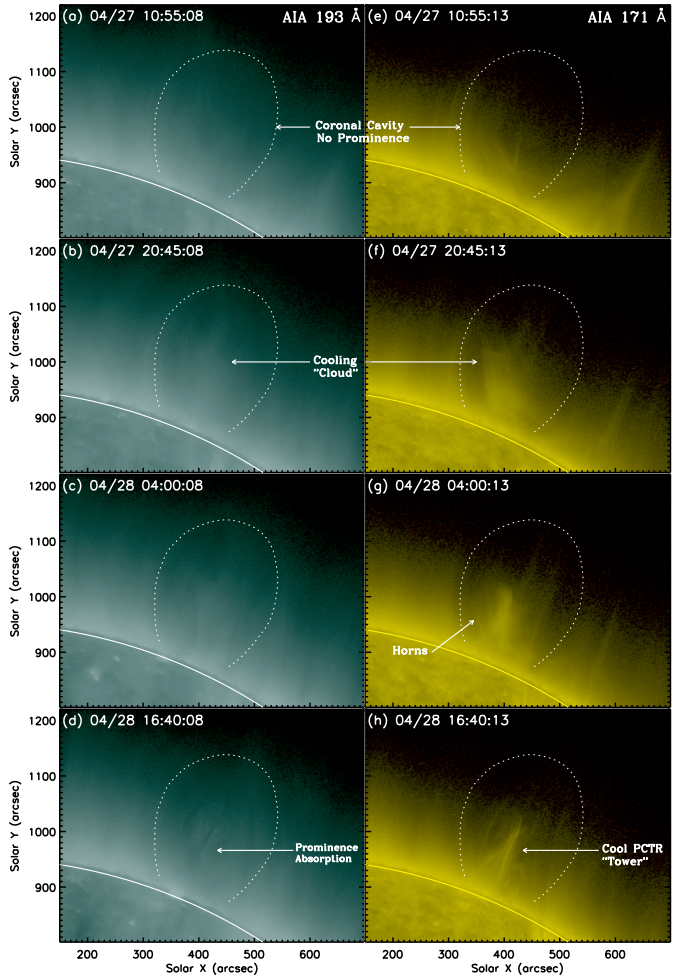} 
	\caption{Comparison of coronal cavity emission cloud appearance in the 193~\AA\ (left) and 171~\AA\ (right) channels over time. Note that the emission cloud fades in the 193~\AA\ channel, (b)--(c), while it brightens in the 171~\AA\ channel, (f)--(g). See also Movie 2.}
	\label{fig:193+171}
\end{figure}

Figure~\ref{fig:193+171} shows a subset of the temporal sequence discussed above in the \ion{Fe}{12} 193~\AA\ and 171~\AA\ channels. The subset begins at 04/27 10:55~UT when there was no prominence visible in the 304~\AA\ channel, and shows a largely featureless, barely visible, coronal cavity in the 193~\AA\ channel while the center of the cavity is well-defined in the 171~\AA\ channel. The second row of Fig.~\ref{fig:193+171} shows the clearly developed cloud structure in both channels.  By 04/28 04:00~UT the cloud has decreased in contrast in the 193~\AA\ channel while remaining bright and high contrast in the 171~\AA\ channel (third row of Fig.~\ref{fig:193+171}).  The fourth row shows that the prominence has fully formed by 04/28 16:40~UT. Absorption of the 193~\AA\ emission by both the prominence and some of the residual horn structure (see Fig.~\ref{fig:171+304}(h)) is evident. 

\begin{figure}
	\includegraphics[width=0.45\textwidth]{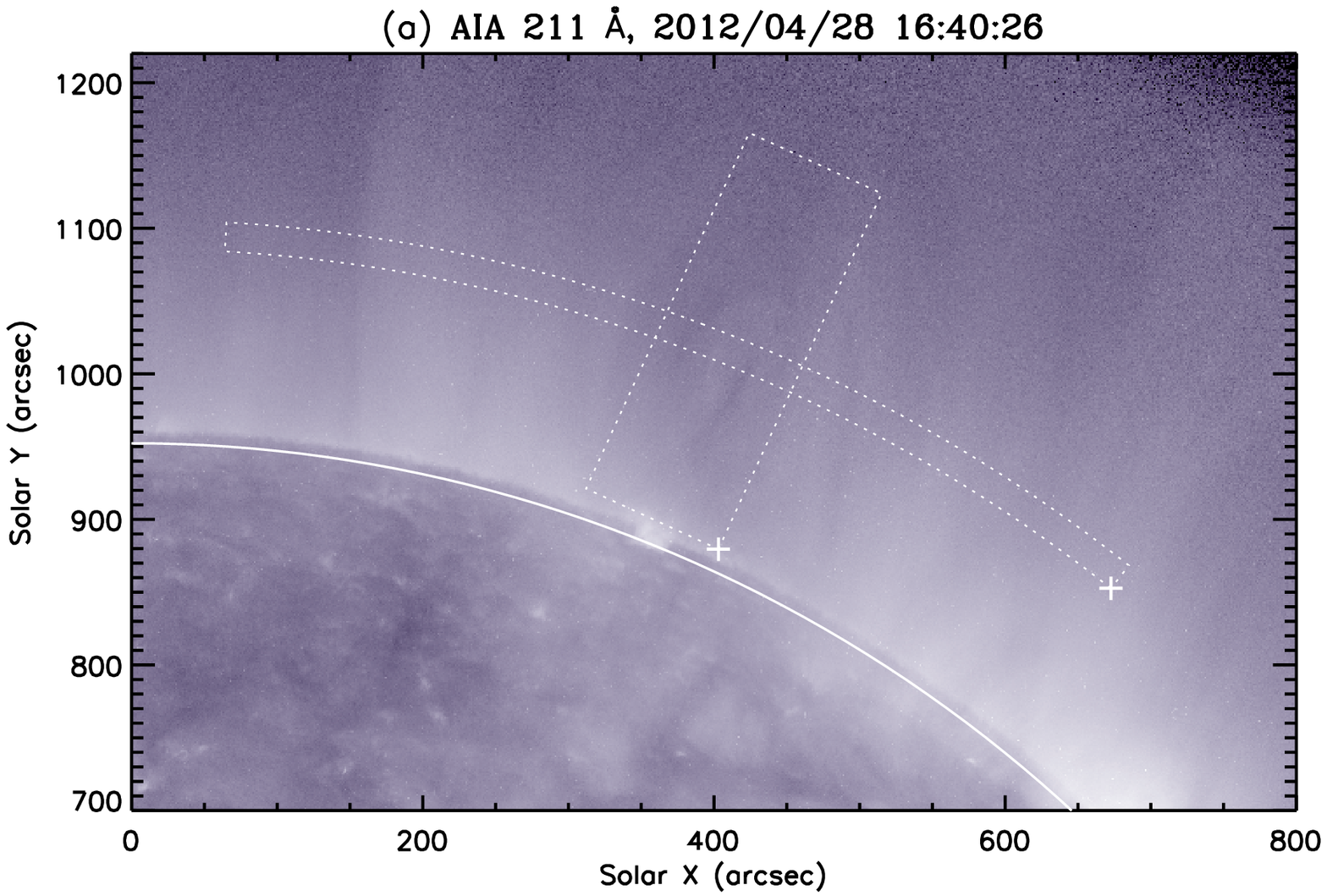} \\ 
	\includegraphics[width=0.45\textwidth]{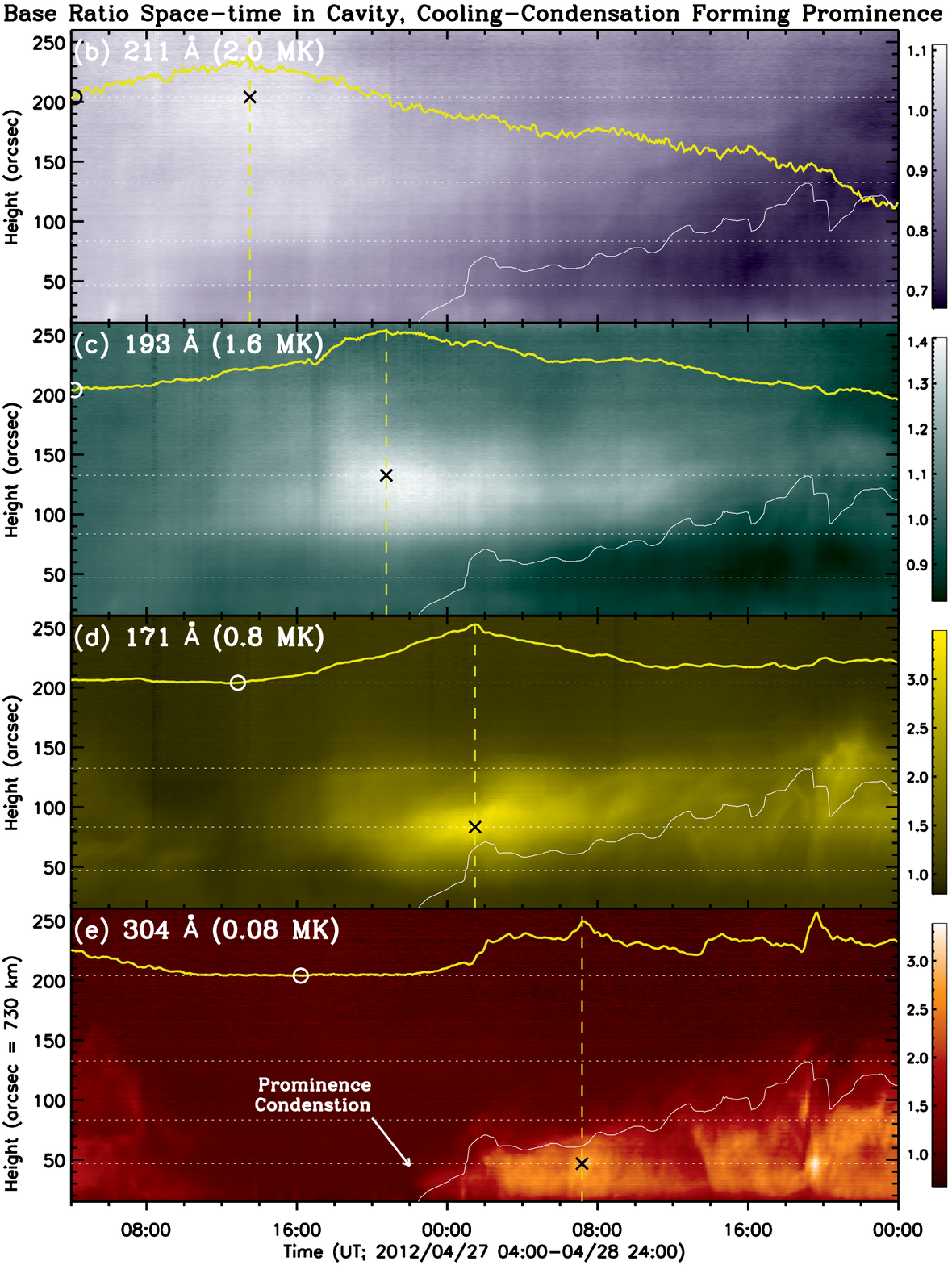} 
	\caption{Timeslice analysis of the emission cloud. (a) 211~\AA\ channel image taken when the prominence has fully formed. The cavity outline is well visible as is the absorption due to the prominence. The vertical dotted box indicates the region over which vertical columns of pixels are averaged to create timeslice images in the 211, 193, 171, and 304~\AA\ channels. The constant height dotted box ($h = 144\arcsec$ above the limb) is used in the timeslice of Fig.~\ref{fig:mass}.  White `\textsf{+}'s mark the lower left corners of the timeslices. (b)--(e):  base ratios with the reference image time for each panel indicated by a small circle. Colorbars indicate the intensity ratio ranges. Black `\textsf{X}'s mark the locations and times of peak emission in each panel. The yellow curves plot relative emission (arbitrarily scaled) along the horizontal dashed lines that traverse the cavity at heights corresponding to the peak emission in each channel. The white contour in all panels shows the outline of the prominence in the 304~\AA\ channel. The peak temperature of emission in each channel is indicated next to the wavelength identifier in units of $10^6$~K (MK).}
	\label{fig:timeslice}
\end{figure}

Figure~\ref{fig:timeslice} shows the emission in the 211, 193, 171, and 304~\AA\ channels averaged over slices taken from the vertical box that encloses the central region of the cavity shown in Panel~(a). Panels~(b)--(e) show the resulting ``timeslices'' in each channel. Emission in the 211~\AA\ channel is spatially concentrated and brightest (relative to the reference frame) at  04/27 13:30~UT. Emission in the 193~\AA\ channel is spatially most compact and brightest several hours later at 20:45~UT. Similarly the 171~\AA\ emission is compact and brightest at 04/28 01:28~UT. Finally, the prominence has its brightest, most compact 304~\AA\ emission at 04/28 07:10~UT. Note that the height of maximum emission steadily decreases with time and decreasing wavelength (and temperature) in the 211, 193, and 171~\AA\ sequence.  The first appearance of the 304~\AA\ prominence emission occurs only after the cloud has descended to a height of approximately 75~Mm in the 171~\AA\ channel. 

\section{Filter Ratio Temperature Analysis}
\label{sect_temp}

\begin{figure}
	\includegraphics[width=0.48\textwidth]{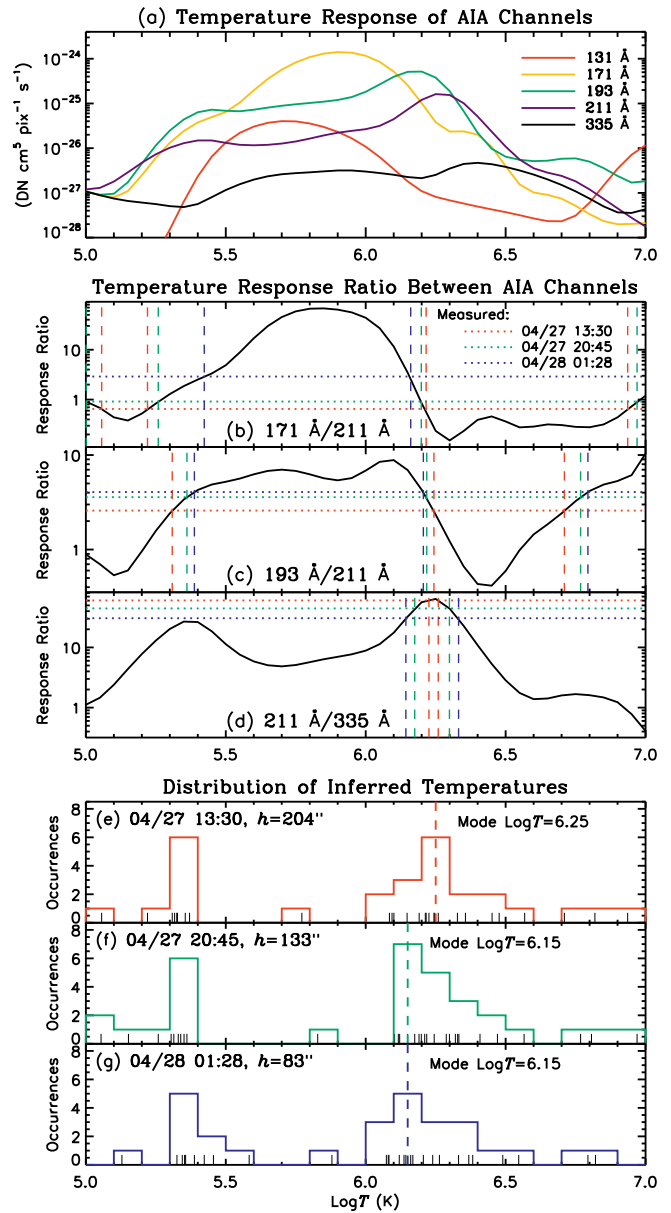} 
	\caption{Filter ratio analysis of emission cloud temperature.  (a) AIA temperature response curves for the five EUV wavelength channels used in the ratio analysis. (b)--(d) Example 171/211, 193/211, and 211/335 response ratio curves. For the three times in the timeslice sequence indicated by \textsf{X}-marks in Figs.~\ref{fig:timeslice}(b)--(d), we plot the corresponding ratios measured in the emission cloud. The colored horizontal dotted lines show the ratio measurements for each time while the vertical dashed lines indicate the intersection points with the ratio response curves. Colors delineate the times as shown in (b).  The histogram of intersection points is shown in (e)--(g) for the three times, respectively. In these panels, all ratio intersection points are shown at each time as short vertical bars (see text). The modal temperature for the hotter peak is indicated in $\log\mathrm{T}$ form in each panel. The histogram colors correspond to the times shown in (b).}
	\label{fig:twofilter}
\end{figure}

Given the wide spectral passbands of AIA, a detailed determination of the plasma temperature and density along the lines of  \cite{Heinzel:2008ig} is not possible. However we can use the known AIA instrument response functions to calculate a temperature range for the emitting cloud structure as a function of time.  While this method implicitly assumes an isothermal temperature distribution for the plasma \citep[which is unlikely to be valid in a complex coronal cavity, e.g.,][]{Kucera:2012}, it has the advantage of having no adjustable parameters and is not an ill-posed inversion problem. 

Figure~\ref{fig:twofilter}(a) shows the measured temperature response curves of the AIA telescopes from \texttt{aia\_get\_response.pro}. Taking the ratios 171/211, 193/211, 211/335  gives the set of curves shown in Fig.~\ref{fig:twofilter}(b)--(d). The emission of the cloud in each channel is measured by averaging the emission of 3$\times$3 ``pixel'' samples around the \textsf{X}-marks shown at specific times in Fig.~\ref{fig:timeslice}(b)--(d)\footnote{The average is thus taken over the three images before, during, and after the peak emission times marked in Fig.~\ref{fig:timeslice}(b)--(d) and in three spatial pixel vertically centered on the location of the `\textsf{X}' marks.}. The ratios are then computed and plotted on Figs.~\ref{fig:twofilter}(b)--(d) as the dotted horizontal lines.

We measured all ten ratios from the set $\{131, 171, 193, 211, 335\}$\AA\ at each of the specific times above. The temperatures at the intersections of these ten ratios with the response curves are shown as short vertical bars on the axes of Figs.~\ref{fig:twofilter}(e)--(g). These measurements are binned in temperature to create histograms of inferred temperatures, showing that there are two most likely solutions around $\log\mathrm{T}\sim 5.35$ and $6.2$~K. The peak around $\log\mathrm{T}\sim 6.2$~K shows a larger degree of clustering than the lower temperature peak implying that it is the more likely solution. In addition, this peak shows a trend toward a cooler temperature over the three time steps sampled in the dataset. In particular, the mode of the peaks are $\log\mathrm{T}\sim 6.25$, $6.15$, and $6.15$~K for the times 04/27 13:30~UT, 04/27 20:45~UT, and 04/28 01:28~UT, respectively. The latter peak, while having the same mode as the previous time, shows a pronounced tail to the cooler temperature range. In contrast, the $\log\mathrm{T}\sim 5.35$~K peak does not show such a consistent trend. 

Taken together, the filter ratio analysis implies that the initial temperature of the emission cloud is $\log\mathrm{T}\sim 6.25$~K with a cooling trend over time. Emission ultimately appears in the 304~\AA\ channel implying a very definite cooling trend, however this filter cannot be used in the foregoing filter ratio analysis because the plasma is optically thick in \ion{He}{2}, violating the optically thin assumption in the response curves of Fig.~\ref{fig:twofilter}.

\section{Prominence Mass and Cavity Contrast Comparison}
\label{sect_mass}

\begin{figure}
	\includegraphics[width=0.5\textwidth]{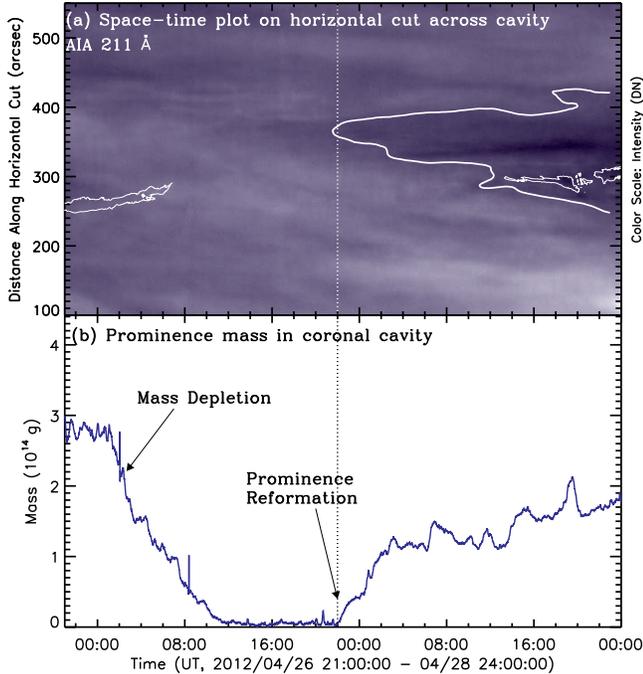} 
	\caption{Coronal cavity contrast and prominence mass cycle.  (a) Timeslice image in the  211~\AA\ channel formed from the average of all constant height pixels in the dotted outline in Fig.~\ref{fig:timeslice}(a).  The thick white outline shows the coronal cavity as it reaches its highest contrast. The darkest central core of the cavity appears at a position of 340\arcsec, offset by approximately 40\arcsec\ from the location of the 304~\AA\ prominence indicated by the thin white contours. (b) 304~\AA\ ``plane-of-sky'' mass of the prominence as a function over the same time period. The POS mass increases roughly simultaneously with the increase in coronal cavity contrast.}
	\label{fig:mass}
\end{figure}

Figure~\ref{fig:mass} shows a contrast analysis of the coronal cavity in the 211~\AA\ channel along with a mass estimate in the quiescent prominence. Cavity contrast is calculated by examining the average emission in a timeslice of constant-height pixels (dotted outline in Fig.~\ref{fig:timeslice}(a)). Figure~\ref{fig:mass}(a) shows the timeslice and reveals that the core of the cavity darkens significantly beginning around 04/27 16:00~UT. This is several hours after the peak 211~\AA\ emission of the cooling cloud, but roughly coincides with the rise to maximum emission in the 193~\AA\ cloud, as shown in Fig.~\ref{fig:timeslice}. As the cloud emission intensifies in the cooler channels, the cavity darkens further until by about 04/28 16:00~UT, the cavity achieves maximum darkness in the core.  Although cavity contrast is a strong function of view angle and hence can be influenced by solar rotation, at the heliographic latitude of the cavity (N68), the rotation is only $6^\circ$ over the 12~h during which the cavity contrast changes significantly, too little to explain the observed contrast change. We note also that the darkest region of cavity extends over a larger region than the prominence absorption. 

We estimate the prominence mass as the ``plane-of-sky'' (POS) mass, calculated by measuring the area of emission detected in the 304~\AA\ channel and assuming that the  prominence  is  a uniform thickness slab in the POS. This method grossly underestimates the true mass of the prominence since most prominences have significant longitudinal extent. Nevertheless, the method allows an analysis of prominence mass over time at a zero-order level of detail; changes in the assumed geometry of the prominence will alter only the absolute mass number and not the shape of the temporal curve. Here we assume a typical prominence density of $\mathrm{n_{e}} = 8\times 10^9 \pcmc$ \citep{Labrosse:2010bt} or $\rho = 1.3\times 10^{-14}\ \mathrm{g} \pcmc$ ,  and a thickness of 10\arcsec\ (7300~km) to arrive at the numbers shown in Fig.~\ref{fig:mass}(b). 

The plot shows that the POS mass of the prominence steadily decreases at an average rate of approximately $2.4\times 10^{13}$~g~h$^{-1}$ from 00:00~UT to 12:00~UT on 27-April. Beginning at about 22:00~UT, the new prominence forms initially at about the same rate. 
However the rate decreases after 04/28 3:00~UT and is highly variable thereafter. Note that the onset of rapid coronal cavity darkening in the 211~\AA\ channel corresponds well with the onset of prominence appearance in the 304~\AA\ channel and that the cavity core increases in darkness as the prominence continues to accumulate mass (see Movie 1b). Since coronal cavity core contrast is largely a function of plasma density relative to the surrounding corona \citep{Fuller:2009jm}, we interpret the darkening of the cavity core to indicate a loss of plasma from the region.

\section{Summary and Discussion}	
\label{sect_discussion}

We have analyzed observations of the dynamic formation of a quiescent polar crown prominence in a coronal cavity. The sequence of events is summarized as follows:
\begin{enumerate}
\item A pre-existing prominence slowly disappears due mostly to drainage and lateral transport of plasma (Fig.~\ref{fig:171+304}).
\item A bright emission cloud forms in the upper regions of the coronal cavity (Figs.~\ref{fig:171+304} and \ref{fig:193+171}).
\item The cloud descends toward the lower region of the cavity while sequentially becoming brighter in the 211, 193, and 171~\AA\ channels (Fig.~\ref{fig:timeslice}).
\item A new prominence appears in the 304~\AA\ channel and rapidly grows in both vertical and horizontal extent (Figs.~\ref{fig:171+304} and \ref{fig:timeslice}).
\item The coronal cavity core above the prominence grows darker in the 211~\AA\ channel as the 304~\AA\ prominence grows (Fig.~\ref{fig:mass}). 
\item When the prominence reaches its maximum size after approximately 18~h of growth, the emission cloud in the cavity is completely gone (Fig.~\ref{fig:171+304}).
\end{enumerate}

Taken together, these observations are consistent with the hypothesis that the quiescent prominence has formed via condensation from hotter plasma contained in the core of the coronal cavity. We interpret the EUV emission sequences in Figs.~\ref{fig:171+304},  \ref{fig:193+171}, and \ref{fig:timeslice} as evidence of radiative cooling and descent of the hot plasma in the coronal cavity core, first appearing bright in the hotter  211~\AA\ channel with a peak temperature of formation of $\log\mathrm{T_{max}}\sim 6.2$~K, followed by a shift to emission in the 193~\AA\ channel ($\log\mathrm{T_{max}}\sim 6.1$~K), followed by another shift to emission in the  171~\AA\ channel ($\log\mathrm{T_{max}}\sim 5.8$~K), and finally appearing in the ``chromospheric'' 304~\AA\ channel ($\log\mathrm{T_{max}}\sim 4.8$~K). 

The profile of the hotter filter ratio histogram widens toward cooler temperatures over time (Fig.~\ref{fig:twofilter}(g)), further supporting our interpretation that the observed plasma radiatively cools to condense into the prominence.  The drop in height of the cloud further supports this hypothesis. Radiative cooling is proportional to $\mathrm{n_{e}}\!^2$ at corona temperatures and increases with decreasing temperature in the $\log\mathrm{T} \sim$ 5--7~K range. Therefore a ``runaway thermal instability'' is possible in which the plasma cools increasingly rapidly as it becomes denser and cooler \citep{Low:2012tu}. The increased density of the cloud naturally results in higher gravitational force countering the local Lorentz force of the cavity magnetic field. Thus the descent of the cloud is consistent with radiative cooling and condensation in a weakly magnetized environment as the plasma seeks a new equilibrium height, possibly through resistive reconnection \citep{Low:2012tv}. As the prominence gains mass at the expense of the hot cloud, the coronal cavity simultaneously becomes more sharply defined in the 211~\AA\ channel due to mass loss from the core region.

The observations shown here imply the possibility of \emph{in situ} formation of prominences in coronal cavities. Given the large density disparity, \cite{TandbergHanssen:1995ud} estimated that the entire coronal mass could only support such condensation for a few large prominences.  However this concern was based on the implicit assumption that prominences are magnetostatic suspensions of a fixed amount of plasma. But as shown by recent SDO/AIA and \emph{Hinode}/SOT observations, quiescent prominences are in constant motion via drainage downflows \citep{Liu:2012gq,Haerendel:2011de,Chae:2008ca} and re-supply of mass through various mechanisms \citep[e.g.,][]{Berger:2011eoa,Li:2012,Su:2012}.  \cite{Liu:2012gq} estimated the mass loss in prominence downflows  to be $10^{15}$~g/day or roughly the equivalent of a typical CME mass. Thus the mass in a prominence changes significantly with time and is not directly related to the mass in the global corona in a simple manner. Mass balance in prominences appears to be a cyclic process in which the mass transported into the coronal cavity determines how much can eventually condense to form prominences. Similar cyclic processes have been proposed for the quiet Sun coronal mass balance \citep{McIntosh:2012,Antolin:2012,Marsch:2008} and coronal rain condensation \citep{Landi:2009,Schrijver:2001}.

Observations at the limb necessarily integrate along heliographic longitude. Thus it is possible that the prominence that disappears is not at the same longitude as the new prominence that forms, or that the cloud and the new prominence are a line-of-sight coincidence. But the fact that the shrinkage of the cloud (darkening of the cavity), the descent of the cloud, the formation of prominence ``horns'', and the eventual formation of a PCTR sheath are all synchronized with the growth of the prominence in the 304~\AA\ channel argue against a line-of-sight coincidence. We note that the hot cloud itself exhibits no significant flows, other than its slow radial descent, during the condensation process. Thus it is unlikely that we are confusing \emph{in situ} condensation with flow-based formation mechanisms such as flows between disparate magnetic footpoints \citep[e.g.,][]{Xia:2012,Luna:2012eb,Karpen:2008hc,Antiochos:1999ef}.

\end{document}